\documentclass[twocolumn,aps,pre,preprintnumbers,amsmath,amssymb,nofootinbib,floatfix]{revtex4-1}

\usepackage{graphicx,color}
\usepackage{caption}
\usepackage{subcaption}
\usepackage{amsmath}
\usepackage{epstopdf}
\usepackage{soul}
\usepackage{tabularx}
\newcommand{\D}{\mathrm{d}}
\newcommand{\e}{\mathrm{e}}
\newcommand{\half}{\frac{1}{2}}
\newcommand{\be}{\begin{equation}}
\newcommand{\ee}{\end{equation}}
\newcommand{\bea}{\begin{eqnarray}}
\newcommand{\eea}{\end{eqnarray}}
\newcommand{\ba} {\begin{align} }
\newcommand{\ea} {\end{align} }
\newcommand{\eps}{\varepsilon}

\newcommand{\kbt}{k_{\mathrm{B}}T}

\newcommand{\lb}{l_{{\rm B}}}

\newcommand{\vecr}{\boldsymbol{r}}
\newcommand{\veck}{\boldsymbol{k}}
\newcommand{\kmax} {k_{\rm max}}
\newcommand{\emf} {\varepsilon_{_{\rm MF}}}

\newcommand {\kmf} {\kappa_{_{\rm MF}}}

\newcommand{\ra}[1]{\textcolor{black}{#1} } 

\begin{document}


\title{Dielectric Constant of Ionic Solutions: \\ Combined Effects of Correlations and Excluded Volume }

\author{Ram M. Adar$^{1}$, Tomer Markovich$^{2}$, Amir Levy$^3$, Henri Orland$^{4,5}$, David Andelman$^1$}
\email{andelman@post.tau.ac.il}
\affiliation{$^1$Raymond and Beverly Sackler School of Physics and Astronomy\\ Tel Aviv
University, Ramat Aviv, Tel Aviv 69978, Israel\\
$^2$DAMTP, Centre for Mathematical Sciences, University of Cambridge, Cambridge CB3 0WA, United Kingdom\\
$^3$Department of Physics, Massachusetts Institute of Technology Cambridge, MA 02139, USA\\
$^4$Institut de Physique Th\'eorique, CE-Saclay CEA, F-91191 Gif-sur-Yvette Cedex, France\\
$^5$Beijing Computational Science Research Center, No.10 East Xibeiwang Road, Beijing 100193, China}


\begin{abstract}
The dielectric constant of ionic solutions is known to reduce with increasing ionic concentrations. However, the origin of this effect has not been thoroughly explored. In this paper we study two such possible sources: long-range Coulombic correlations and solvent excluded-volume.
Correlations originate from fluctuations of the electrostatic potential beyond the mean-field Poisson-Boltzmann theory, evaluated by employing a field-theoretical loop expansion of the free energy. The solvent excluded-volume, on the other hand, stems from the finite ion size, accounted for via a lattice-gas model. We show that both correlations and excluded volume are required in order to capture the important features of the dielectric behavior. For highly polar solvents, such as water, the dielectric constant is given by the product of the solvent volume fraction and a concentration-dependent susceptibility per volume fraction. The available solvent volume decreases as function of ionic strength due the increasing volume fraction of ions. A similar decrease occurs for the susceptibility  due to correlations between the ions and solvent, reducing the dielectric response even further. Our predictions for the dielectric constant fit well with experiments for a wide range of concentrations for different salts in different temperatures, using a single fit parameter related to the ion size.
\end{abstract}

\maketitle

\section{Introduction}
\label{sec1}
Ionic solutions are ubiquitous in electrochemical, colloidal, and biological systems, and are most commonly studied via the Poisson-Boltzmann (PB) theory. Within this mean-field (MF) theory, ions are treated as point-like and interact solely via the Coulomb interaction, while the solvent is modeled as a homogeneous dielectric medium~\cite{David95,Israelachvili}. Although rather simplistic, the PB theory captures the important underlying physical principles of charged objects in solutions, and is in good agreement with experiments for weakly charged surfaces and monovalent salts in low concentrations.

The PB theory has nevertheless several limitations. As a MF theory, it neglects correlations between ions and fluctuations of the electrostatic potential. These corrections to MF are especially important~\cite{Abrashkin07,Netz00,Moreira00,Henderson79,Nielba85} for multivalent ions and strongly charged surfaces, colloids, polyelectrolytes, and other charged macromolecules. The finite size of ions also plays a significant role, giving rise to short-range steric interactions and limiting ionic concentrations by their close packing value. Such finite-size effects can be included in the PB theory in what is known as the modified PB (MPB) theory~\cite{Borukhov97,Borukhov00}. The van der Waals (vdW) interaction, neglected in PB theory, plays an important role as well. Incorporating vdW interactions with PB theory leads to the well-known Derjaguin-Landau-Verwey-Overbeek (DLVO) theory, which successfully explains the stability of colloidal suspensions~\cite{VO}. Finally, molecular dynamics (MD) simulations that incorporate corrections to DLVO have been used in recent years to study  specific models for solvent and solute molecules~\cite{Patey75,Sharma07,Chandra00,Kalcher09,Chowdhuri01,Zhu92,Azuara08}.

PB theory treats all ions on the same footing and neglects ionic specific effects~\cite{Kunz10,Levin,Levin1,Tomer14,Dan11,Dan11a}. Such an important ionic specific effect is the dielectric decrement of ionic solutions~\cite{HastedBook,DebyeBook,BarthelBook,Hasted48,Fawcett95} as function of the ionic concentration.  The overall change in the dielectric constant of an ionic solution can be large and reduce the dielectric constant by $50\%$. It leads to significant differences in the behavior of ionic solutions near interfaces and surfaces~\cite{Kunz10,Dan11,Dan11a} and affects the ionic activity coefficient~\cite{Vincze10}.

For dilute ionic solutions (usually  $n<1$\,M), it is observed in experiments that the dielectric constant decreases linearly with the ionic concentration, {\it i.e.},
\be
\label{eq1}
\eps\left(n\right)=\eps_w-\gamma n,
\ee
where $\eps(n)$ is the dielectric constant of the solution, $n$  the ionic concentration, $\eps_w$  the dielectric constant of pure solvent (usually water), and $\gamma$ is a positive constant, measured in M$^{-1}$. Experiments show that $\gamma$ is ionic specific and roughly ranges from $5\,{\rm M}^{-1}$ to $22\,{\rm M}^{-1}$ \cite{Hasted48,Fawcett95} for simple cations and anions. The linear dependence can be interpreted in terms of hydration shells. Each ion in the solution interacts strongly with the surrounding polar solvent that forms a hydration shell. The solvent molecules in the shell are not as free to rotate and align in response to an external field as those far from the ion. This results in a dielectric decrement when a dilute ionic solution is considered.  The dielectric response is affected further by ions due to their polarizability~\cite{Demery12}, but as we consider in this work simple and small ions, this effect will be neglected hereafter.

At higher ionic concentrations, the dielectric decrement is no longer linear~\cite{Friedman82,Amir,Amir1,Gavish16,Kjellander,Kjellander1}. Friedman~\cite{Friedman82} analyzed a site-site interaction model and described the decrement in terms of  molecular direct correlation functions and second moments of site-site correlation functions. Gavish and Promislow~\cite{Gavish16}  considered the local electric field that ions exert on the solvent at high ionic concentrations, and wrote the dielectric function in terms of the excess ionic polarization and molten salt dielectric constant. Kjellander~\cite{Kjellander,Kjellander1} described the dielectric response of dressed ions and solvent with effective dipolar (and higher) moments, due to electrostatic correlations.

In this work, we revisit the problem of the dielectric decrement and derive an improved analytical expression for the dielectric constant of ionic solutions. This expression is written in terms of a single physical quantity related to the ionic size. We demonstrate how excluded volume and electrostatic correlations beyond MF affect the dielectric constant. The former effect is incorporated via a lattice-gas model, while the latter are captured using a field-theoretical loop expansion of the free energy. For simplicity, other non-Coulombic interactions ({\it e.g.}, hydrogen bonds), dynamic effects~\cite{Wolynes79,Wolynes80}, and  ionic contributions to the dielectric constant due to ionic polarizability~\cite{Demery12} and ion pairs~\cite{Bjerrum,Zwanikken09,Adar17} are neglected.

The outline of this paper is as follows. In Sec.~\ref{sec2}, we describe our microscopic lattice-gas model and formulate the loop expansion of the free energy. In Sec.~\ref{sec3}, a general expression for the dielectric constant is derived, and the contribution of fluctuations beyond MF is highlighted. We then focus in Sec.~\ref{sec4} on aqueous solutions with a high dielectric constant. We compare the relative contributions of excluded volume and electrostatic correlations to the dielectric decrement, and expand the dielectric function as function of the ionic concentration for low concentrations. In Sec.~\ref{sec5}, our predictions are compared to experimental data and are shown to be in very good agreement. Finally, in Sec.~\ref{sec6}, we provide some general observations and concluding remarks.

\section{Model}
\label{sec2}

Consider an aqueous solution with monovalent ions of bulk concentration $n$. We describe the solution as a lattice gas, and divide it into cubic cells of dimensions $a\times a\times a$. Each cell can be occupied by either a cation (charge $e$) or an anion (charge $-e$). Here we assume that the two ionic species occupy a similar volume in the solution. Cells vacant of ions are occupied by a solvent with a dipolar moment ${\bf p}$.
As cells can be occupied by a single ion at most, the lattice-gas model accounts for steric effects between the different species, inducing a short-range repulsive interaction.

We write the partition function in terms of spin-like
variables, following the derivation of Ref.~\cite{Borukhov00}. Each
cell $j,$ located at position $\boldsymbol{r}_{j},$ is described by a variable, $s_{j}$, with $s_j=\pm1$ for a cell occupied by a cation or anion, respectively, and $s_j=0$ for a cell occupied by the solvent. With these variables, the charge density operator is given by
\begin{align}
\label{eq2}
\widehat{\rho}(\vecr) & =\rho_{f}(\vecr)+\sum_{j}\left[es_j-\left(1-s_j^2\right)p\hat{\boldsymbol{n}}_j\cdot\boldsymbol{\nabla}\right]\delta\left(\vecr-\vecr_{j}\right),
\end{align}
where $\delta(\vecr)$ is the Dirac delta function, and $\rho_{f}(\vecr)$ is a possible fixed (immobile) charge density. The first term within the sum accounts for occupancy by a cation or an anion, while the second term in the sum corresponds to a solvent dipole in the $j$-cell with a dipole moment $\boldsymbol{p}_j=p\hat{\boldsymbol{n}}_j$.

It is possible to write the partition function, $\Xi$, in terms of the charge density operator, $\widehat{\rho}$~\cite{Borukhov00,Amir,Amir1},
\begin{align}
\label{eq3}
\Xi & =\sum_{s_{j}}\prod_{j}\e^{\beta\mu s_{j}^2}\int\frac{\D\Omega_{j}}{4\pi}\nonumber \\
&\times\exp\left[-\frac{\beta}{2}\int\int \D^3r\,\D^3r'\,\widehat{\rho}\left(\boldsymbol{r}\right)v_{c}\left(|\boldsymbol{r}-\boldsymbol{r'}|\right)\widehat{\rho}\left(\boldsymbol{r}'\right)\right].
\end{align}
In Eq.~(\ref{eq3}), $\mu=\mu_\pm$ is the ionic chemical potential, which is equal for both positive and negative ions due to electroneutrality, $\beta=1/\left(k_{B}T\right)$ is the inverse thermal energy, $\Omega_{j}$ is the solid angle of $\hat{\boldsymbol{n}}_j$, and $v_{c}\left({r}\right)=1/\left(4\pi\eps_{0}r\right)$
is the Coulomb interaction kernel in SI units.

We replace the lattice-occupation degrees of freedom, $\{s_j\}$, with an auxiliary field by introducing a density field, $\rho\left(\boldsymbol{r}\right)$,
and its conjugate field, $\varphi\left(\boldsymbol{r}\right),$ via
the functional identity:
\begin{align}
\label{eq4}
1 & =\int\mathcal{D}\rho\,\delta\left[\rho\left(\boldsymbol{r}\right)-\widehat{\rho}\left(\boldsymbol{r}\right)\right]\nonumber \\
 & =\int\mathcal{D}\rho\mathcal{D}\varphi\exp\left(i\beta\int \D^3r\varphi\left(\boldsymbol{r}\right)\left[\rho\left(\boldsymbol{r}\right)-\widehat{\rho}\left(\boldsymbol{r}\right)\right]\right),
\end{align}
where $\int\mathcal{D}\rho$ denotes a functional integral over the field $\rho$, and similarly $\int\mathcal{D}\varphi$ for $\varphi$. Substituting Eq.~(\ref{eq4}) in Eq.~(\ref{eq3}) ultimately leads to the functional integral form~\cite{Abrashkin07},
\begin{align}
\label{eq5}
\Xi & =\mathcal{N}\,\int\mathcal{D}\varphi\,\e^{-\beta S\left[\varphi\right]}\,,
\end{align}
where $\mathcal{N}$ is a non important prefactor and $S$ is the field action
\begin{align}
\label{eq6}
S\left[\varphi\right] & =\int \D^3r\left[\frac{\eps_{0}}{2}\left(\nabla\varphi(\vecr)\right)^{2}+i\rho_{f}(\vecr)\varphi(\vecr)\right.\nonumber \\
 &\left. -\frac{1}{\beta a^{3}}\ln\big[{\rm sinc}\left(\beta p\left|\boldsymbol{\nabla}\varphi(\vecr)\right|\right)+2\Lambda \cos\left(\beta e \varphi(\vecr)\right)\big]\right].
\end{align}
\ra{In the above equation, we make use of the function ${\rm sinc}(x)=\sin x / x$} and the ionic fugacity, $\Lambda=\exp\left(\beta \mu\right)$.  Note that a sum over the discrete lattice sites is replaced in Eq.~(\ref{eq6}) by an integral over space, as is appropriate in the continuum limit, resulting in the $1/a^{3}$ factor.

The free energy is related to the partition function via
\be
\label{eq7}
F=-\kbt\ln\Xi,
\ee
and the bulk ionic concentration, $n$, is related to the fugacity, $\Lambda$, via
\be
\label{eq8}
n=-\frac{\Lambda}{2V}\frac{\partial\beta F}{\partial \Lambda},
\ee
where $V$ is the total volume. Note that $n$ is the bulk concentration of both cations and anions, {\it i.e.}, $n_+=n_-=n$, resulting in the factor of two in Eq.~(\ref{eq8}). Finally, a relation exists between the auxiliary field, $\varphi$, and electrostatic potential, $\psi$. Applying the identity $\psi=\delta F/\delta \rho_f$ to Eqs.~(\ref{eq5})-(\ref{eq7}), we find that
\be
\label{eq9}
\psi=\frac{1}{\Xi}\int\mathcal{D}\varphi \,\,i\varphi\,\e^{-\beta S[\varphi]}\equiv  i\langle \varphi\rangle,
\ee
{\it i.e.}, up to the imaginary unit, the electrostatic potential is the thermal average of the auxiliary field, $\varphi$.

\subsection{The loop expansion }
The partition function of Eq.~(\ref{eq5}) is written as a functional integral. This integral cannot be performed analytically, but can be calculated within some approximation. We employ a systematic saddle-point expansion~\cite{Schwinger} of the partition function, referred to in Quantum Field Theory (QFT) as a {\it loop expansion}. The partition function at the saddle point yields the MF theory, as is described below, and the Gaussian fluctuations around the saddle point result in the {\it one-loop} correction term.

It is convenient to introduce (in analogy with QFT) an artificial expansion parameter that multiplies the field action, $S\to \ell^{-1} S$, and will be set to unity at the end. The parameter $\ell$ plays the role of $\hbar$ in QFT, and is useful in order to keep track of orders in the expansion.
The saddle point value of the field $\varphi$ is denoted as $\varphi_0$, and the second functional derivative of the field action, evaluated at $\varphi_0$, as
\be
\label{eq10}
\left.S_2\left(\vecr,\vecr'\right)=\frac{\delta^2S\left[\varphi\right]}{\delta\varphi(\vecr)\delta\varphi(\vecr')}\right|_{\varphi=\varphi_0}.
\ee
The loop expansion of the free energy then reads
\be
\label{eq12}
F\approx S\left[\varphi_0\right]+\frac{\ell}{2\beta}{\rm Tr}\left(\ln\beta S_2[\varphi_0]\right),
\ee
where the logarithm of an irrelevant prefactor is omitted.

 In order to perform a consistent calculation, all physical quantities are expanded up to first order in $\ell$. For example, the fugacity is written as $\Lambda=\Lambda_{\rm MF}+\ell\Lambda_1$, where $\Lambda_{\rm MF}$ is the MF fugacity and $\Lambda_1$ is the one-loop correction. The terms $\Lambda_{\rm MF}$ and $\Lambda_1$  are found by solving Eq.~(\ref{eq8}) consistently up to first order in $\ell$. The MF value is given by
\be
\label{eq13}
\Lambda_{\rm MF}=\half\frac{\Phi}{1-\Phi},
\ee
where $\Phi=2n a^3$ is the volume fraction occupied by the two ionic species in the bulk.

\subsection{The MDPB equation}
The loop expansion is performed around the MF value of the auxiliary field, $\varphi$, as determined by the Euler-Lagrange equation for the free-energy functional, $F$. This is a MF equation and constitutes a generalized PB equation, named the {\it Modified Dipolar PB} (MDPB) equation. It contains two added effects: (i) steric modifications of the standard PB theory as in MPB, and, (ii) inclusion of dipoles, as in dipolar PB theory (DPB)~(see Eqs.~(8) and~(9) of Ref.~\cite{Abrashkin07}).

On the MF level, $\varphi=-i\psi$ and $\Lambda$ is given by Eq.~(\ref{eq13}). The MDPB equation then reads
\begin{align}
\label{eq14}
\eps_0\nabla^2\psi&=-\rho_f+2 n e \frac{\sinh\left(\beta e \psi\right)}{D_{\rm MF}}\nonumber\\
&+\nabla\cdot\left[\left(a^{-3}-2n\right)\frac{g\left(\beta pE\right)}{D_{\rm MF}}\mathcal{L}\left(\beta pE\right)p\hat{\boldsymbol{E}}\right],
\end{align}
where we have denoted $\boldsymbol{E}=E\hat{\boldsymbol{E}}=-\nabla\psi$ as the electric field, $g(u)=\sinh u/u$ and $\mathcal{L}(u)=\coth u-1/u$ is the Langevin function. The MF denominator in Eq.~(\ref{eq14}), $D_{\rm MF}$, is given by the weighted average:
\be
\label{eq15}
D_{\rm MF}=\Phi\cosh\left(\beta e\psi\right)+\left(1-\Phi\right)g\left(\beta p E\right).
\ee

We address the physical origins of the terms on the right-hand-side (RHS) of Eq.~(\ref{eq14}). The first line accounts for the charge-density, consisting of any fixed charges and the mobile ions. Compared to the standard PB theory, the ionic charge density is divided by the denominator function, $D_{\rm MF}$. This function leads to a saturation of the ionic and dipolar concentrations, bounded from above by the close-packing density, $a^{-3}$~\cite{Borukhov97,Borukhov00}. The second line in Eq.~(\ref{eq14}) accounts for the divergence of the polarization density, written as the product of the dipole density, $\left(a^{-3}-2n\right)g\left(\beta pE\right)/D_{\rm MF}$, and the average dipole moment, $\mathcal{L}\left(\beta pE\right)p\hat{\boldsymbol{E}}$.

In this paper, we are concerned with calculating the dielectric constant in the bulk electrolyte, far away from any immobile charged objects. Hence, we set hereafter a zero density of any fixed charges, $\rho_f=0$. The solution to the MDPB equation in this case is simply $\psi=0$.

\subsection{One-loop free energy}
Substituting the field action of Eq.~(\ref{eq6}) in Eq.~(\ref{eq12}), and replacing $\varphi$ by the electrostatic potential $\psi$ according to Eq.~(\ref{eq9}), we find the following expression for the free energy:
\begin{align}
\label{eq16}
F[\psi]& =\int \D^3r\bigg[-\frac{\eps_{0}}{2}E^2(\vecr)-\frac{1}{\beta a^{3}}\ln\big[2\Lambda \cosh\left(\beta e \psi(\vecr)\right)\nonumber\\
&+g\left(\beta pE(\vecr)\right)\big]\bigg]+\frac{\ell}{2\beta}{\rm Tr}\ln \left(\beta S_2\right),
\end{align}
where $S_2$, the second functional derivative of the action, is given by
%
\begin{align}
\label{eq17}
S_2&=-\eps_0\nabla^2\delta\left(\vecr-\vecr'\right)\nonumber\\
&+\frac{4\pi\eps_0 l_0}{a^3 D}\left(\left[2\Lambda\cosh\left(\beta e\psi\right)-\frac{4\Lambda^2\sinh^2\left(\beta e \psi\right)}{D}\right]\delta\left(\vecr-\vecr'\right)\right.\nonumber\\
&+4\Lambda b\nabla\cdot\left(\frac{\sinh\left(\beta e \psi\right)}{D}\int\frac{\D^2\Omega}{4\pi}\,\boldsymbol{\hat{ n}}\e^{-\beta p\boldsymbol{\hat{n}}\cdot\boldsymbol{E}}\right)\delta\left(\vecr-\vecr'\right)\nonumber\\
&-b^2\nabla\cdot\left[\left(\int\frac{\D^2\Omega}{4\pi}\,\boldsymbol{\hat{n}\hat{n}}'\e^{-\beta p\boldsymbol{\hat{n}}\cdot\boldsymbol{E}}\right.\right.\nonumber\\
&\left.\left.\left.-\frac{1}{D}\int\frac{\D^2\Omega}{4\pi}\frac{\D^2\Omega'}{4\pi}\,\boldsymbol{\hat{n}\hat{n}}'\e^{-\beta p\left(\boldsymbol{\hat{n}}+\boldsymbol{\hat{n}}'\right)\cdot\boldsymbol{E}}\right)\cdot\nabla\delta\left(\vecr-\vecr'\right)\right]\right).
\end{align}
In Eq.~(\ref{eq17})  we have denoted $l_0= e^2/\left(4\pi\eps_0\kbt\right)$ as the vacuum Bjerrum length, to be distinguished from the Bjerrum length in solution with $\lb=e^2/\left(4\pi\eps_w\kbt\right)$, and $b=p/e$ is a typical length scale of the solvent dipole. The unit vectors $\boldsymbol{\hat{n}}$ and $\boldsymbol{\hat{n}}'$ point in the direction of the solid angles $\Omega$ and $\Omega'$, respectively, which are integrated over.   The denominator function, $D$, is given by
\begin{align}
  \label{eq17b}
D=2\Lambda \cosh\left(\beta e \psi\right)+g\left(\beta pE\right).
    \end{align}
Note that the MF denominator function of Eq.~(\ref{eq15}), $D_{\rm MF}$, is related to $D$ via $D_{\rm MF}=\left.\left(1-\Phi\right) D\right|_{\Lambda=\Lambda_{\rm MF}}.$

The dielectric constant can be obtained from the free energy via~\cite{Amir,Amir1}
\begin{align}
\label{eq18}
\eps&=\left.-\int\D^3r\,\frac{\delta^2 F}{\delta E_i(\vecr)\delta E_i(0)}\right|_{\psi=0},
\end{align}
where $E_i=-\partial\psi/\partial r_i$ is the $i$th component of the electric field, and any of $i=1,\,2,\,3$ can be equally used for isotropic systems, such as the one discussed.

\section{Results}
\label{sec3}
The dielectric constant is obtained by applying Eq.~(\ref{eq18}) to the one-loop free energy of Eq.~(\ref{eq16}). The dielectric constant is written according to the loop expansion as $\eps=\emf+\ell\eps_1$, where $\emf$ is the MF dielectric constant and $\eps_1$ is the one-loop correction term. The two terms are presented and discussed in length below.

\subsection{MF dielectric constant}

Retaining zeroth order terms in $\ell$, both in the free energy of Eq.~(\ref{eq16}) and the fugacities, leads to the MF dielectric constant
\be
\label{eq19}
\frac{\emf}{\eps_0}=\left(1-\Phi\right)\delta,
\ee
where we have denoted $\delta$ as a dimensionless parameter for the solvent susceptibility,
\be
\label{eq19b}
\ra{\delta= \frac{1}{3}\frac{p^2}{\eps_0\kbt a^3}.}
\ee
 It quantifies the electrostatic energy of solvent dipoles within a unit cell in terms of the thermal energy. For the sake of clarity, in Table~\ref{table1} we distinguish between several quantities used to describe the dielectric constant. The expression of Eq.~(\ref{eq19}) is typical of effective medium theory, where the contribution of each species is weighted by its volume fraction in the solution.

The result of Eq.~(\ref{eq19}) yields the linear decrement coefficient $\gamma$ of Eq.~(\ref{eq1}) on the MF level, $\gamma_{\rm MF}$ (where $\gamma=\gamma_{\rm MF}+\ell\gamma_1$). We find that the MF dielectric constant, $\emf$,  decreases linearly with the ionic concentration with the coefficient
\be
\label{eq20}
\gamma_{\rm MF}=2a^3\delta.
\ee
The decrement originates solely from the finite ionic size. The volume of each ion in the solution comes at the expense of the polar solvent and lowers the dielectric response.

The electrostatic interaction between ions and solvent dipoles also lowers the dielectric constant. Dipoles become oriented towards ions or away from them, and are less free to rearrange and align in response to an external electric field. Such correlations are captured in the one-loop correction below, and result in a modified $\gamma$ coefficient. Furthermore, at high ionic concentrations, correlations lead to a non-linear dielectric decrement.

For pure solvent, Eq.~(\ref{eq17}) gives $\emf=\eps_{0}\left(1+\delta\right)$. This is a known MF result~\cite{Abrashkin07} for a dilute phase of dipoles of concentration $a^{-3}$, and does not produce the correct dielectric constant of pure water, where dipoles interact quite strongly with one another. For example, substituting  $a=3.1$\,\AA, corresponding to a concentration  of $55\,$M, the value $\eps_w\simeq78\,\eps_0$ is obtained for $p_{w}\simeq4.8\,{\rm D}$, more than twice as large as the physical value $p_{w}=1.85\,{\rm D}$ of water molecules. This discrepancy stems from solvent dipole-dipole correlations that largely determine the dielectric response and are not accounted for on this MF level. This issue is resolved in the one-loop correction, as is described below. Note that water molecules interact also via other non-electrostatic interactions, such as {\it hydrogen bonds} that modify the dielectric response. However, such interactions lie beyond the scope of this paper.

\begin{table}[ht]
\begin{tabular}{l l}
   \hline
  $\eps_0$ & vacuum permittivity   \\
   $\delta$ &  solvent susceptibility scale (dimensionless)  \\
  $\emf$ & MF dielectric constant \\
  $\eps_1$ & one-loop correction to the dielectric constant \\
  $\eps$& total dielectric constant of the solution\\
  $\eps_w$ & dielectric constant of pure solvent \\ \hline
  \end{tabular}
  \caption{Different parameters used in relation to the dielectric constant.}
  \label{table1}
  \end{table}

\subsection{One-loop correction}
The one-loop correction to the dielectric constant is obtained by retaining the first-order terms in the free energy and fugacity. The calculation involves logarithmic derivatives of the one-loop free energy term. The logarithmic derivatives of any operator $\mathcal{O}$ involve inverse operators, according to
\be
\label{eq21}
\frac{\partial {\rm Tr} \left[\ln \mathcal{O} \right]}{\partial\alpha}=\int \D^3r\,\int \D^3r'\,\mathcal{O}^{-1}\left(\vecr,\vecr';\alpha\right)\frac{\partial \mathcal{O}\left(\vecr,\vecr';\alpha\right)}{\partial \alpha},
\ee
where $\alpha$ is an arbitrary parameter. The relevant operator for our calculation is $S_2$, as was defined in Eq.~(\ref{eq10}). The inverse operator of $S_2$ is the two-point Green's function~\cite{Tomer16}, $G$, and for $\psi=0$, it is given by
\be
\label{eq22}
S_2^{-1}\left(\vecr,\vecr'\right)\equiv G\left(\vecr,\vecr'\right)=\frac{\kbt}{4\pi\eps_{{\rm MF}}}\frac{e^{-\kmf |\vecr-\vecr'|}}{|\vecr-\vecr'|}.
\ee
In Eq.~(\ref{eq22}), $\kmf$ is the inverse screening length, evaluated on the MF level,
\be
\label{eq23}
\kmf=\sqrt{\frac{2n e^2}{\emf\kbt}}.
\ee
Note that $G$ has units of inverse length and that $1/\kmf$ is different from the classical Debye screening length,  obtained by replacing $\emf$ with the pure water (solvent) value, $\eps_w$, in Eq.~(\ref{eq23}).

Following the definition of Eq.~(\ref{eq18}), we find the following expression for the dielectric constant (for a detailed calculation, see Appendix),
\begin{align}
\label{eq24}
\frac{\eps}{\eps_0}&=1+\frac{\delta}{1+2\Lambda}\nonumber\\
&+\frac{\delta}{\left(1+2\Lambda\right)^2}\left(4\pi l_0\Lambda G(0)-\frac{1-3\Lambda}{3} a^3\delta\nabla^2G(0)\right)\ell.
\end{align}
The result above is written in terms of the full fugacity, $\Lambda$. For consistency, it should be further expanded to first order in $\ell$ with $\Lambda=\Lambda_{\rm MF}+\ell\Lambda_1$. The MF fugacity is given by Eq.~(\ref{eq13}),  and the one-loop correction is found from Eq.~(\ref{eq8}) to be
\be
\label{eq25}
\Lambda_1=\frac{1}{4}\frac{\Phi}{1-\Phi}\left[4\pi l_0 G(0)+a^3\delta \nabla^2G(0)\right].
\ee

Both Eqs.~(\ref{eq24}) and (\ref{eq25}) are written in terms of the Green's function and its Laplacian at the origin, $r\to0$, where they diverge. We surpass these divergences by introducing a cutoff length as the minimal possible distance between the particles (or the maximal wavelength, $\kmax$). Within our model, this minimal distance can be conveniently identified with the lattice constant $a$. Accordingly, we approximate the Green's function and its Laplacian at the origin as
\begin{align}
\label{eq26}
G(0)&=\int_{|\veck|<\kmax}^{}\frac{\D^3k}{\left(2\pi\right)^3}\widetilde{G}(\veck),\nonumber\\
\nabla^2G(0)&=-\int_{|\veck|<\kmax}^{}\frac{\D^3k}{\left(2\pi\right)^3}k^2\,\widetilde{G}(\veck),
\end{align}
where $\widetilde{G}(\veck)$ is the Fourier transform of $G(\vecr)$ and $\kmax=2\pi/a$. Substituting Eq.~(\ref{eq22}) in Eq.~(\ref{eq26}) leads to
\begin{align}
\label{eq27}
G(0)&=\frac{1}{2\pi^{2}\emf}\left(\kmax-\kmf\arctan\frac{\kmax}{\kmf}\right),\nonumber\\
\nabla^2G(0)&=\kmf^2G(0)-\frac{\kmax^3}{6\pi^{2}\emf}.
\end{align}

Inserting the one-loop fugacity of Eq.~(\ref{eq25}) in Eq.~(\ref{eq24}) and expanding in powers of $\ell$, we find the following one-loop correction term in the dielectric constant:
\begin{align}
\label{eq28}
\frac{\eps_1}{\eps_0}=&-\frac{1}{3}\left(1-\Phi\right)^2a^3\delta^2\nabla^2G(0).
\end{align}
The one-loop correction to the dielectric constant is, therefore, quadratic in the solvent volume fraction~\cite{foot}, $1-\Phi$. As the Green's function in Eq.~(\ref{eq28}) is by itself a nonlinear function of the ionic concentration [Eq.~(\ref{eq27})], the dependence of Eq.~(\ref{eq28}) on $\Phi$ is nonlinear as well.

The overall dielectric constant is given by the sum of Eqs.~(\ref{eq19}) and (\ref{eq28}) as
\be
\label{eq28b}
\frac{\eps}{\eps_0}=1+\left(1-\Phi\right)\delta-\frac{1}{3}\left(1-\Phi\right)^2a^3\delta^2\nabla^2G(0),
\ee
where we have set $\ell=1$. This expression for the solution dielectric constant, $\eps$, is our main result. In what follows, we simplify Eq.~(\ref{eq28b}) further by relating $\delta$ and $a$. Then, in Sec.~\ref{sec4}, we analyze in detail an approximation of Eq.~(\ref{eq28b}) for highly polar solvents, such as water.

\subsection{Adjustment of the solvent dipolar moment}
The result of Eq.~(\ref{eq28b}) depends on the two model parameters: the lattice constant, $a$, and dipolar moment, $p$. However, $a$ and $p$ are related via the pure solvent dielectric constant, $\eps_w$.  Next, we describe how to adjust the solvent dipolar moment for given $\eps_w$ and $a$ values. With this procedure, it is possible express the results only in terms of $\eps_w$ and $a$.

In order to understand the relation between $p$ and $a$, we consider two containers with two different electrolyte solutions. The solutions share the same solvent but consist of monovalent ions of different sizes. The first solution is modeled by a lattice constant $a$, and the second by $a'$ that is larger than $a$. The two solutions are then diluted into pure solvent. As a result of this process, the same solvent is described differently in the two containers. In the first, it is described by cells of volume $a^3$ with a dipolar moment $p$, while in the second, by cells of volume $a'^3$ with a dipolar moment $p'$.

Clearly, the two descriptions above must yield the same experimentally observed solvent dielectric constant, {\it i.e.},
\be
\label{eq29b}
\ra{\eps\left(\Phi=0,a,p\right)=\eps\left(\Phi=0,a',p'\right) = \eps_w.}
\ee
This constraint is most conveniently expressed via the solvent susceptibility scale, $\delta$. Inserting $\Phi=0$ in Eq.~(\ref{eq28b}) results in
\be
\label{eq29}
\frac{\eps_w}{\eps_0}=1+\delta+\frac{4\pi}{9}\frac{\delta^2}{1+\delta}.
\ee
Inverting the above equation leads to
\begin{align}
\label{eq29a}
\delta&=\frac{9}{2\left(9+4\pi\right)}\Bigg[\frac{\eps_w}{\eps_0}-2\nonumber\\
&+\sqrt{\left(\frac{\eps_w}{\eps_0}-2\right)^2+4\left(\frac{\eps_w}{\eps_0}-1\right)\left(1+\frac{4\pi}{9}\right)}\,\,\Bigg],
\end{align}
where only the positive root of Eq.~(\ref{eq29}) was chosen. Therefore, a given $\eps_w$ corresponds to a single $\delta$ value, leading to $p\propto a^{3/2}$. In terms of our example above, this result implies that a larger unit cell  has a larger dipolar moment, {\it i.e.},  $p'>p$.

It is possible to substitute Eq.~(\ref{eq29a}) for $\delta$ in Eq.~(\ref{eq28b}) to obtain the dependence of the dielectric constant on the ionic concentration in terms of a single model parameter, $a$. Inserting the dipole moment of water $p=1.8$\,D and matching to $\eps=78\,\eps_0$, yields $a=2.15\,{\rm \AA}$.  This value is comparable with the diameter of water, which is considered to be about $2.7\,{\rm \AA}$~\cite{Svishchev93}.

\section{Simplified $\eps$ for highly polar solvents}
\label{sec4}

In what follows, we consider highly polar solvents with $\eps_w\gg\eps_0$, such as water with $\eps_w\simeq78\,\eps_0$. For such high $\eps_w$ values, Eq.~(\ref{eq29a}) can be approximated according to
\be
\label{eq29b}
\delta=\frac{9}{9+4\pi}\frac{\eps_w}{\eps_0}\simeq0.42\frac{\eps_w}{\eps_0}.
\ee
Using the above value of $\delta$ in the $\emf$ expression of Eq.~(\ref{eq19}) for $\Phi=0$, it is evident that the one-loop solvent-solvent correlation term is responsible for over a half of the pure solvent dielectric response. Substituting Eq.~(\ref{eq29b}) in Eq.~(\ref{eq28b}) while omitting zeroth-order terms in $\delta$ ($\delta\gg1$ for $\eps_w\gg\eps_0$), yields a compact expression for the dielectric constant,
\be
\label{eq34}
\frac{\eps}{\eps_w}=\left(1-\Phi\right)\left[1-c\left(\frac{\kmf a}{2\pi}\right)\right],
\ee
where
\be
\label{eq35}
c(x)=\frac{12\pi}{4\pi+9}\left(x^2-x^3\arctan\frac{1}{x}\right).
\ee
Equation~(\ref{eq34}) elucidates the combined effect of excluded volume and correlations. The dielectric response is given by the product of the solvent volume fraction in the solution, $1-\Phi$, and an effective relative solvent susceptibility, $1-c$.

While excluded volume is a purely steric effect, the effective susceptibility of the solvent depends on both excluded volume and correlations. This is evident from the argument $x=\kmf a/\left(2\pi\right)$ of the function $c(x)$, which can be written as
\be
\label{eq36}
\frac{\kmf a}{2\pi}=\left(\frac{9+4\pi}{9\pi}\right)^{1/2}\sqrt{\frac{\Phi}{1-\Phi}\frac{\lb}{a}}\simeq 0.87\sqrt{\frac{\Phi}{1-\Phi}\frac{\lb}{a}}.
\ee
The expression above demonstrates that the significance of electrostatics is determined by the ratio $\lb/a$. Steric effects, on the other hand, depend on the ratio $\Phi/\left(1-\Phi\right)$.

\subsection{Correlations vs. excluded volume}
\label{ssec4a}

Equation~(\ref{eq34}) highlights the relative contributions of the two mechanisms behind the dielectric decrement: excluded volume and electrostatic correlations. We compare between the two by examining the derivative of Eq.~(\ref{eq34}) with respect to $\Phi$,
\be
\label{eq36a}
\frac{1}{\eps_w}\frac{\D \eps}{\D \Phi}=-\left(1-c\right)-\left(1-\Phi\right)\frac{\D c}{\D \Phi}.
\ee
The expression above has a simple physical interpretation. When the ion volume fraction is increased by a small amount, $\D \Phi$, the same amount is excluded from the solvent. As the contribution per solvent volume fraction is $1-c$, this results in a decrement of $\left(1-c\right)\D\Phi$. This is the first term on the RHS of Eq.~(\ref{eq36a}).

At the same time, the contribution per volume fraction of the remaining solvent depends on the ionic concentration due to electrostatic correlations, and changes by an amount $-\left(\D c/\D\Phi\right)\D\Phi$. Multiplying by the solvent volume fraction results in a decrement of $\left(1-\Phi\right)\left(\D c/\D\Phi\right)\D\Phi$. This is the remaining term on the RHS of Eq.~(\ref{eq36a}).

\begin{figure}[ht]
\centering
\includegraphics[width=0.85\columnwidth]{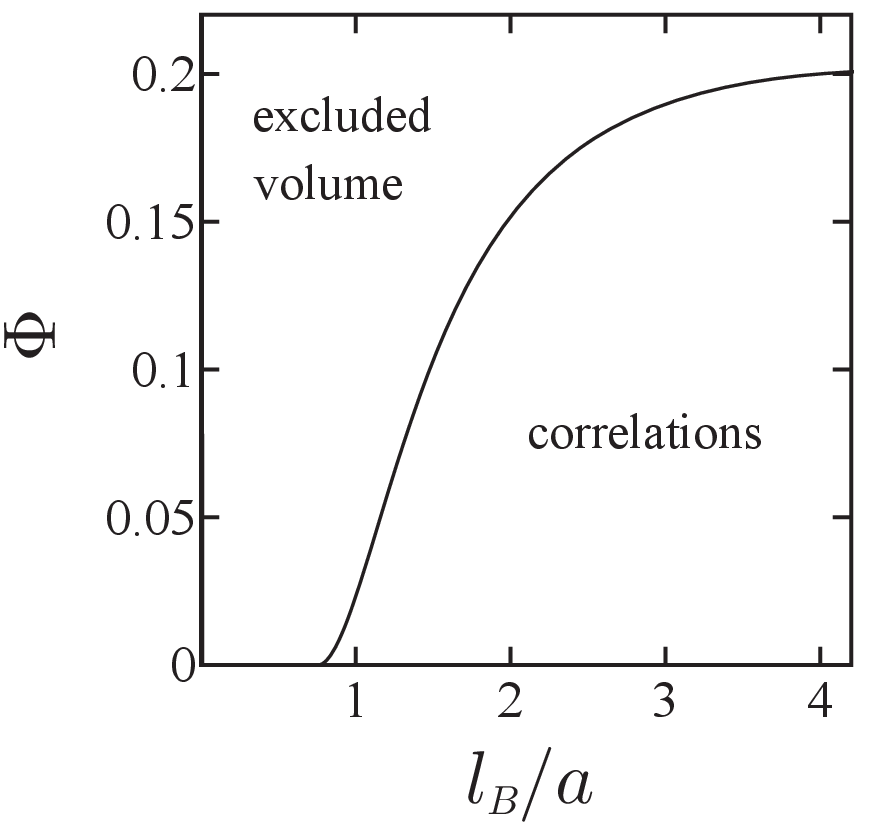}
\caption{Two dielectric decrement regimes in the $\left(\lb/a,\Phi\right)$ plane. The crossover line is given by Eq.~(\ref{eq36b}) and separates the regime dominated by excluded volume on top from that dominated by correlations on the bottom.
}
\label{fig1}
\end{figure}

We equate the two terms described above to determine when excluded  volume and electrostatic correlations have equal contributions to the dielectric decrement. This criterion is given by
\be
\label{eq36b}
\frac{xc'(x)}{1-c(x)}=2\Phi,
\ee
where $c'(x)=\D c/\D x$ denotes a derivative of $c$ with respect to $x=\kmf a/\left(2\pi\right)$. In order to obtain Eq.~(\ref{eq36b}), the relation $\D x/\D \Phi=x/\left[2\Phi\left(1-\Phi\right)\right]$ was used. As $x$ depends on $\Phi$ and on $\lb/a$ [Eq.~(\ref{eq36})], the equality of Eq.~(\ref{eq36b}) relates $\Phi$ and $\lb/a$ values. The crossover line corresponding to this criterion is plotted in Fig.~\ref{fig1} and separates the two dielectric decrement regimes. For $\Phi$ values above the contour, the dielectric decrement is dominated by excluded volume, while for values below it, correlations dominate.

It is evident from Fig.~\ref{fig1} that for small values of $\lb/a\ll1$, excluded volume is dominant  for arbitrarily small $\Phi$ values. However, for more physical values of $\lb/a\simeq2$, as is relevant for Cl$^-$ and F$^-$ ions in aqueous solutions at room temperature (see also Sec.~\ref{sec5}), correlations are dominant for $\Phi\lesssim0.15$. For $a=3.5\,{\rm \AA} \simeq 0.5\,\lb$, this value corresponds to a concentration of $n\simeq5.8$\,M. As all the experimental dielectric data reviewed in this work lie within this range $n<5.8$\,M, they are dominated by correlations.

We emphasize that even when correlations are dominant, excluded volume still plays an important role. In order to determine when the excluded volume effect is negligible, we re-examine Eq.~(\ref{eq36b}) and replace the factor of two by a much larger factor of $20$. This corresponds to $\lb/a$ and $\Phi$ values for which the excluded-volume contribution to the dielectric decrement is ten times smaller than the correlation one. Such a condition can be satisfied for a non-negligible $\Phi$ value only for  $\lb/a\gtrsim10$, not shown in Fig.~\ref{fig1}. For such high $\lb/a$ values, we expect higher orders in the loop expansion to be significant.

In the opposite limit of high volume fractions of ions, $\Phi\to 1$, the excluded volume is always dominant. As we neglect the ionic polarizability and possible ion pairs, the dielectric constant simply vanishes at this limiting value, as is implied by Eq.~(\ref{eq34}). We do not focus on such high concentrations, where the solution may become saturated. Similarly to the case of high $\lb/a$ values, higher-order corrections in the loop expansion become important for this $\Phi\to1$ limit.

\subsection{Low concentration expansion}
\label{ssec4b}
For low ionic concentrations, we expand Eq.~(\ref{eq34}) in powers of $n$, according to
\be
\label{eq32a}
\eps=\eps_w-\gamma\, n+\zeta \,n^{3/2}.
\ee
Expansions of the dielectric constant in such order terms of $n$ are widely used~\cite{BarthelBook}. Note that the first nonlinear correction term is of power $3/2$. This power is recognizable from classical DH theory, where the correction to the ideal gas free-energy is proportional to $n^{3/2}$.

According to Eq.~(\ref{eq34}), the linear decrement coefficient is given by
\be
\label{eq32}
\gamma=2a^3\eps_w\left(1+\frac{4}{3}\frac{\lb}{a}\right).
\ee
The expression above relates the single microscopic parameter in our model, $a$, to the linear dielectric decrement at low ionic concentrations. The first term in Eq.~(\ref{eq32}) is proportional to the volume of a unit cell and describes the purely steric effect (see Sec.~\ref{sec3}). \ra{Hence, the linear coefficient, $\gamma$, is ionic-specific, as was mentioned in Sec.~\ref{sec1}.} Note that the relative contributions of the MF and one-loop terms to the steric decrement are the same as their relative contributions to the pure dielectric constant. The second term in Eq.~(\ref{eq32}) is proportional to the surface area of a unit cell and originates from correlations.

Expanding Eq.~(\ref{eq34}) to next order in the concentration, we find
\be
\label{eq37}
\zeta=\frac{4\eps_w}{9}\sqrt{2\pi\left(9+4\pi\right)}\,\lb^{3/2}a^3\simeq 5.17\eps_w\lb^{3/2}a^3.
\ee
The results for $\gamma$ and $\zeta$, Eqs.~(\ref{eq32}) and (\ref{eq37}), respectively, enable to approximate the concentration $n_\ast$, at which the dielectric decrement deviates from its linear regime in a noticeable way. We define $n_\ast$ as the concentration for which the value of the $n^{3/2}$ term reaches $10\%$ of the linear term, {\it i.e.}, $\zeta \,(n_\ast)^{3/2}/\gamma\, n_\ast=0.1$. This chosen criterion results in
\be
\label{eq38}
n_\ast\simeq 0.0015\left(1+\frac{4}{3}\frac{\lb}{a}\right)^2\lb^{-3}.
\ee

\begin{figure}[ht]
\centering
\includegraphics[width=0.85\columnwidth]{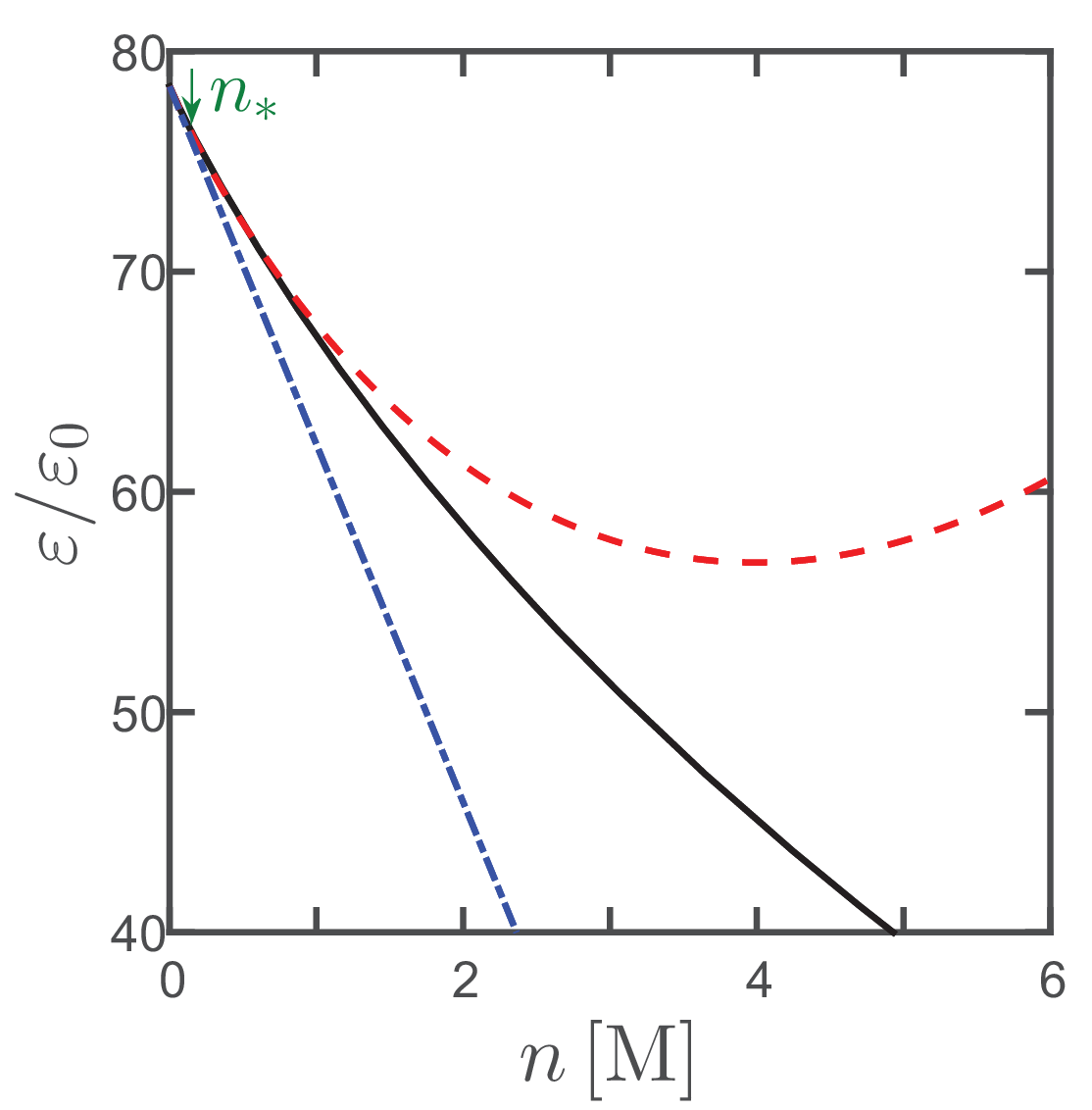}
\caption{(Color online) Dielectric constant as function of ionic concentration, $n$ (in molar), for $\eps_w=78$ and $a=3.6\,{\rm \AA}$, according to  the full one-loop result of Eq.~(\ref{eq34}) (solid black line),  the linear approximation of Eq.~(\ref{eq32}) (dash-dotted blue line), and the leading correction of Eq.~(\ref{eq37}) (dashed red line). The deviation from the linear approximation at $n_\ast\simeq0.094\,{\rm M}$ is marked with by a green arrow, in accordance with Eq.~(\ref{eq38}).
}
\label{fig2}
\end{figure}

For example, for an aqueous NaCl solution at room temperature, we use $\lb=7\,{\rm \AA}$ and $a=3.6\,{\rm \AA}$ (see also Sec.~\ref{sec5}), to obtain $n_\ast\approx0.094\,{\rm M}$ from Eq.~(\ref{eq38}). The deviation from the linear approximation of Eq.~(\ref{eq32}), which corresponds to this $n_\ast$ value, is marked by an arrow in Fig.~\ref{fig2}.
Note that the deviation from the linear decrement is evident for concentrations much lower than $1$\,M.

\section{Comparison to experiments}
\label{sec5}

The static dielectric constant of an aqueous solution cannot
be measured directly. However, it can be extracted from high frequency dielectric data. The frequency dependent
dielectric response, $\eps(\omega)$ is a complex function, which can be
approximated by the Cole-Cole~\cite{Cole} expression,
\be
\label{eq39}
\eps(\omega)=\eps_\infty+\frac{\eps_s-\eps_\infty}{1+\left(i\omega\tau\right)^{1-\alpha}}-i\frac{\sigma_{\rm dc}}{\eps_0\omega},
\ee
where $\eps_\infty$ is the dielectric constant in the high frequency limit, $\eps_s$ is the static dielectric constant (that
is of interest to us), $\tau$ is the dielectric relaxation time, $\alpha$ is the relaxation time distribution parameter, and $\sigma_{\rm dc}$ is the DC conductivity. In the experiments we review here \cite{Wei92,Buchner99,Loginova06}, the dielectric response was measured in frequencies ranging from $45\,{\rm MHz}$ to $25\,{\rm GHz}$, and $\eps_s$ was obtained as a fitting parameter from Eq.~(\ref{eq39}).

\begin{figure}[ht]
\centering
\includegraphics[width=0.85\columnwidth]{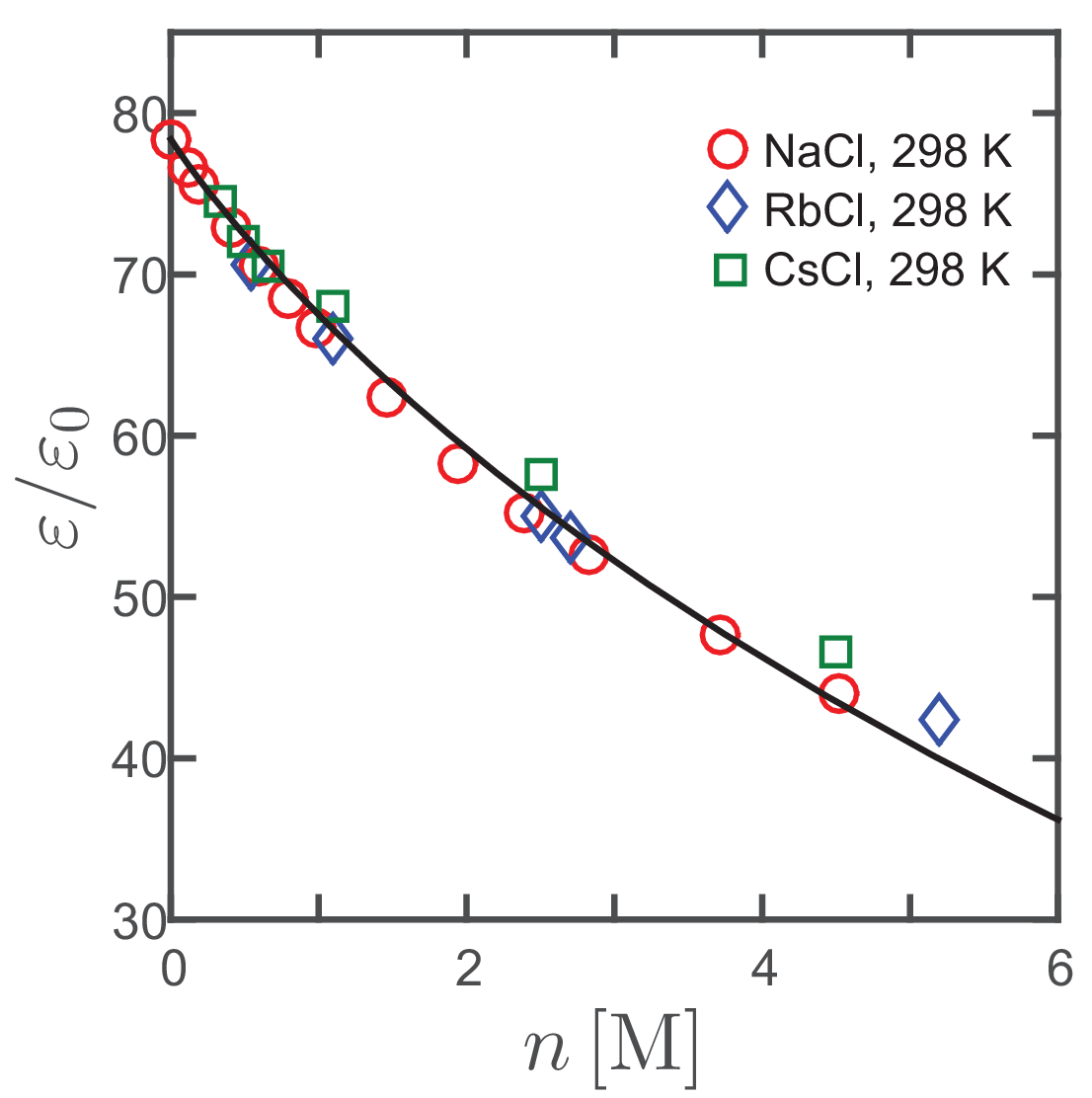}
\caption{(Color online) Comparison of the dielectric constant of Eqs.~(\ref{eq28b}) and (\ref{eq29a})  with experimental data for the static dielectric constant for Cl$^-$  salt solutions. The NaCl data is taken from Ref.~\cite{Buchner99}, and RbCl and CsCl data are taken from Ref.~\cite{Wei92}. The solid curve is plotted by adjusting the only fit parameter,  $a=3.55\,{\rm \AA}$.}
\label{fig3}
\end{figure}

\begin{figure}[ht]
\centering
\includegraphics[width=0.85\columnwidth]{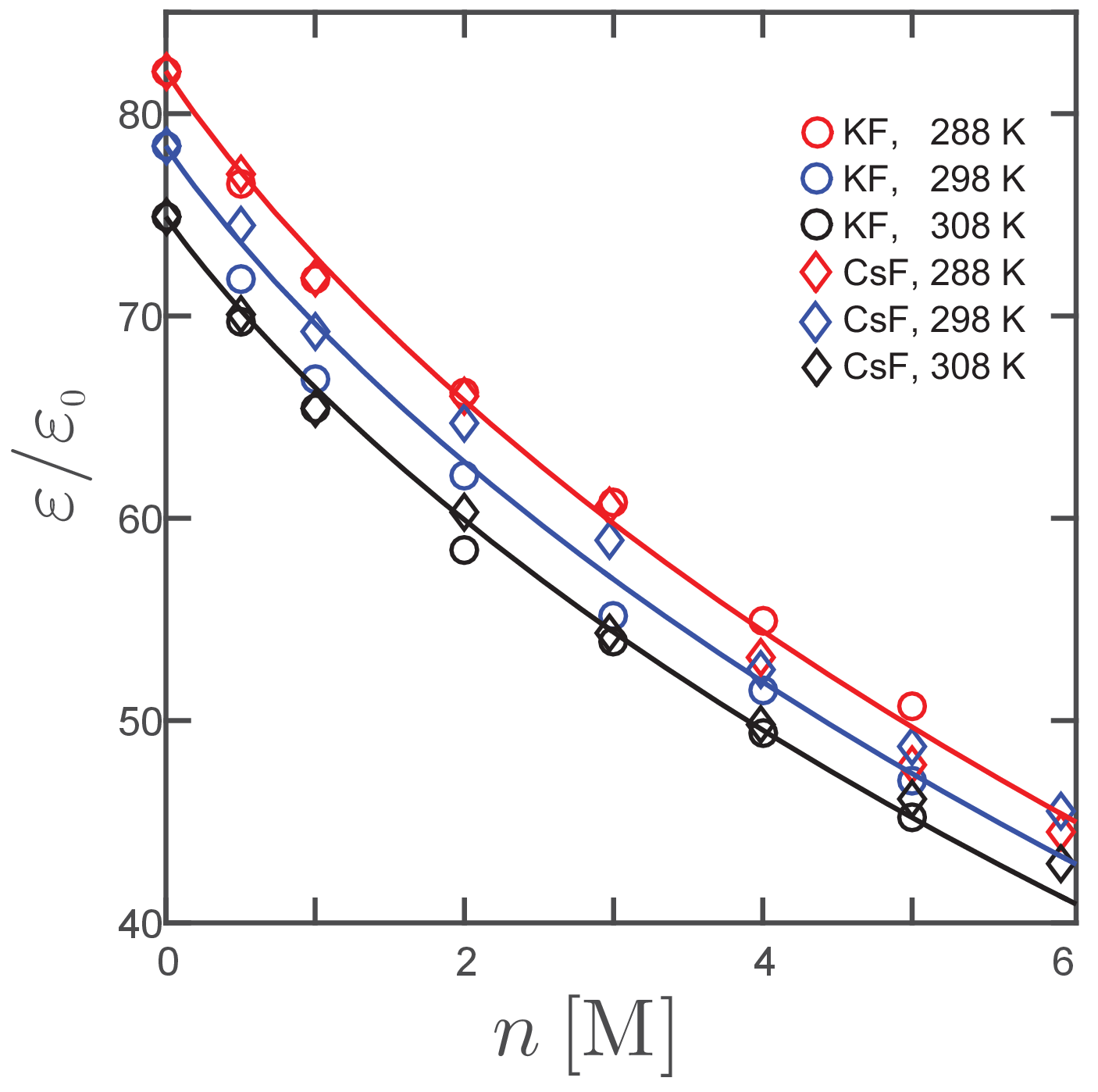}
\caption{(Color online) Comparison of the dielectric constant of Eqs.~(\ref{eq28b}) and (\ref{eq29a})  with experimental data for the static dielectric constant for F$^-$ salt solutions. The data is taken from Ref.~\cite{Loginova06}, and the solid curve is plotted by adjusting the only fit parameter, $a=3.2\,{\rm \AA}$.
 }
\label{fig4}
\end{figure}

We compare our analytical prediction for the dielectric constant, Eqs.~(\ref{eq28b}) and (\ref{eq29a}), to the experimental values of $\eps_s$. The comparison is done for five different ionic solutions, in a wide concentration range of $0-–6\,{\rm M}$ and temperatures that vary between $288$ and $308\,{\rm K}$. We separate the salts into two groups according to their anionic species. Results for three Cl$^-$ solutions at $T=298\,{\rm K}$ are given in Fig.~\ref{fig3}, and those for two F$^-$ solutions at three temperatures, $T=288,$ $298,$ and $308\,{\rm K}$ are given in Fig.~\ref{fig4}. In both figures, $a$ is the only free parameter that is used to fit all the data points. It represents an effective ionic diameter, as will be discussed below.

Our results fit well the experimental data for different salt and temperatures throughout the wide range of concentrations.  In addition, the approximate form of Eq.~(\ref{eq34}) for $\eps_w\gg\eps_0$ can also be used and produces equally adequate fits. From Fig.~\ref{fig3}, it is evident that a single value, $a=3.55\,{\rm \AA}$, is suitable for the three homologous chloride salts, NaCl, RbCl, and CsCl. This value is only slightly smaller than the bare ionic diameter of $\rm{Cl}^{-}$, which is $3.62\,{\rm \AA}$~\cite{Nightingale59}. For all three homologous salts, the Cl$^-$ anion is larger than their cation counterpart. This suggests that $a$ corresponds to the diameter of the largest ion in the solution. Note that the data points for CsCl are slightly higher than the analytical curve, and fitting them separately yields a similar value of $a=3.36\,{\rm \AA}$.

In Fig.~\ref{fig4}, we show that Eqs.~(\ref{eq28b}) and (\ref{eq29a}) with the fit value $a=3.2\,{\rm \AA}$ are in very good agreement with both KF and CsF solutions at three different temperatures. This value is larger than the ionic diameter of F$^{-}$, which is $2.72\,{\rm \AA}$~\cite{Nightingale59}. This discrepancy possibly originates from the specific interaction between the solvent and F$^-$ ions, including the effect of ions on hydrogen bonds (not at all considered here).
Such details lie beyond the scope of this model and are expected to become less important for larger anions and higher temperatures.


\section{Discussion}
\label{sec6}

In this work, we derived an analytical expression for the dielectric constant of ionic solutions. In addition to the ionic concentration, our results depend on the dielectric constant of the pure solvent, $\eps_w$,  Bjerrum length, $\lb$, and lattice constant, $a$. The former quantities are usually known for a given experimental setup, while the $a$ parameter can be extracted by fitting dielectric response data. Our results are in very good agreement with experimental data for different salts in a temperature range of $T=288-308\,{\rm K}$ and salt concentrations as high as $6\,{\rm M}$.

It is shown that a single value of the fit parameter, $a$,  successfully describes several homologous salts at different temperatures. Therefore, once its value is extracted from a given solution for a certain temperature, our theory can be used to predict the dielectric response of many electrolytes  without any further fit parameters. For Cl$^-$ solutions, the value of $a$ is in very good agreement with the bare diameter of $\rm{Cl}^-$. Therefore, it is plausible that for larger anions, one can use our theory without any fit parameters.

We emphasize that $a$ defines the model's minimal length scale, which enters the theory in two ways. First, it defines the lattice constant and, consequently, the excluded volume of solvent. Second, it defines the electrostatic interaction cutoff and, consequently, determines the magnitude of electrostatic correlations. Both the solvent excluded volume and electrostatic correlations are necessary in order to retain the dielectric decrement.
It was shown that for the experimental data discussed in Sec.~\ref{sec5}, correlations play the leading role in the decrement. However, excluded volume is still important and cannot be neglected.

The minimal length scale in our theory \ra{is determined by the size of the ions in the solution. It is} required to be larger than the largest ion diameter, but not necessarily equal to it. A comparison to experimental results (Sec.~\ref{sec5}) indicates that the value of $a$ is indeed mostly determined by the largest of the two species. For anions and cations of the same size, $a$ can be interpreted as the ionic diameter. However, most cation/anion pairs have different sizes, and the exact relation of $a$  to the two ionic diameters is not as clear.

Although $a$ is determined by \ra{the properties of the solute in the solution}, as is discussed above, it is used in our model to describe the solvent as well. Each unit cell of volume $a^3$ that is vacant of ions is occupied by a point-like dipole of moment of moment $p$. In order to consistently describe the solvent and match its pure dielectric constant, we adjust $p$ according to the cell size. This procedure results in  $p\sim a^{3/2}$. Namely, $p$ is not the dipolar moment of a solvent molecule, but rather an effective dipolar moment of solvent within a typical ionic volume. By such a simplified description, we are able to discuss the dielectric constant of aqueous solutions without explicitly addressing the true water molecular charge distribution or hydrogen bonds.

Moreover, solvent-solute interaction is ionic-specific due to the ionic size, the structure of water molecules, and the nature of hydrogen bonds. It is possible to test how ionic specific the dielectric decrement  within our theory is, by fitting it to the measured dielectric data for each electrolyte separately. The best fitting values for  Cl$^{-}$ solutions of Fig.~\ref{fig3} are $a=3.36$\,\AA\, for CsCl, $a=3.5$\,\AA\, for RbCl, and $a=3.62$\,\AA\, for NaCl. The best fitting values for the F$^{-}$ solutions of Fig.~\ref{fig4} at room temperature are $a=3.10$\,\AA\, for CsF and $a=3.22$\,\AA\, for KF. Evidently, deviations from the mean values presented in Figs.~\ref{fig3} and \ref{fig4} are smaller than $10\%$ and may depend strongly on the uncertainty in the experimental data.

Nevertheless, we mention an interesting property of this $a$-dependence. The bare ionic diameters~\cite{Nightingale59} satisfy F$^{-}<$\,Cl$^{-}$ and Na$^{+}<$\,K$^{+}<$\,Rb$^{+}<$\,Cs$^{+}$. Reviewing the values above, we notice that the best-fitted $a$ is larger for larger anions and smaller cations, following the series of the ionic size, but in an opposite manner for cations and anions. Therefore, our results for the dielectric decrement are ionic-specific and follow the Hofmeister series~\cite{Kunz10,Levin,Levin1,Tomer14}. This statement is worth investigating further, but should be considered with caution due to the small magnitude of the effect and possible dependence on uncertainties in the dielectric data.

Another ionic specific property of electrolytes is the tendency of cations and anions to associate into dimers. Such dimers are referred to as {\it Bjerrum pairs}~\cite{Bjerrum,Zwanikken09,Adar17} and contribute further to the electrolyte susceptibility. In Ref.~\cite{Adar17}, a lattice-gas model was proposed to describe such pairs on the MF level, using a phenomenological association energy and permanent pair dipolar moment.  It was shown that the association of ions into pairs can lead to a nonlinear dielectric decrement of aqueous solutions within the MF level (as opposed to this work, where the nonlinear decrement arises only in the one-loop level).

As a side track of the present study, we extended (not shown here) the model of Ref.~\cite{Adar17} to the one-loop level. In the absence of pairs, it reduces exactly to the model presented in this work. It entails a cumbersome one-loop calculation, which  will be presented as part of a future publication~\cite{BP1loop}. For the purposes of the present work, we mention that fitting the augmented model to the experimental data as in Sec.~\ref{sec5}, yields such a large association energy and vanishing pair dipolar moment, that any ion pairing is negligible. Furthermore, the fits produced by such an augmented model were not as good as those of Figs.~\ref{fig3} and \ref{fig4}. We conclude that the theory presented in this work is sufficient in order to describe the dielectric constant of simple aqueous solutions for $n<6$\,M.

However, pairs can play a role for solvents with a low dielectric constant~\cite{BarthelBook}, or specific aqueous solutions at higher concentrations (for example, concentrated LiCl, as was considered in Ref.~\cite{Wei90}). These cases will be explored elsewhere~\cite{BP1loop}.

\vskip 0.5cm
{\it Acknowledgments.~}
We thank R. Colby and S. Safran for fruitful discussions and suggestions. This work was supported
in part by the ISF-NSFC (Israel-China) joint research
program under Grant No. 885/15. D.A. thanks Alexander von Humboldt Foundation for a Humboldt research award, and T.M. acknowledges the Blavatnik postdoctoral fellowship programme at Cambridge University, UK.

\appendix*
\section{Calculation of the one-loop dielectric constant}
\label{appA}

The calculation of the mean field (MF) contribution to the dielectric constant is straightforward, and we focus here on the one-loop correction term. We rewrite the operator $S_2$ for $\psi=0$ in terms of the electric field $\boldsymbol{E}=-\nabla \psi$ and its Cartesian components, $E_j=-\partial\psi/\partial r_j$, as
\begin{align}
\label{eqa1}
S_2\left(\vecr,\vecr'\right)&=\eps_0\left[-\nabla^2\delta\left(\vecr-\vecr'\right)+\frac{4\pi l_0}{a^3}\frac{h(E)}{D(E)}\right],
\end{align}
where we have defined two functions
\begin{align}
\label{eqa2}
D(E)&=2\Lambda+\int \frac{\D^2\Omega}{4\pi}\,\e^{-\beta p_k E_k}\nonumber\\
h(E)&=2\Lambda\delta\left(\vecr-\vecr'\right)+b^2\partial_l\left[\left(-\int \frac{\D^2\Omega}{4\pi} \, \hat{n}_l \hat{n}_m \e^{-\beta p_k E_k}\right.\right.\nonumber\\
&+\left.\left.\frac{1}{D(E)}\int \frac{\D^2\Omega}{4\pi} \, \hat{n}_l \e^{-\beta p_k E_k}\int \frac{\D^2\Omega'}{4\pi}\, \hat{n}'_m \e^{-\beta p_k E_k}\right)\right.\nonumber\\
&\times \partial_m\delta\left(\vecr-\vecr'\right)\Big].
\end{align}
As was defined earlier, $l_0=e^2/\left(4\pi\eps_0\kbt\right)$ is the vacuum Bjerrum length, and $b=p/e$ is a typical length of the solvent dipole. In Eq.~(\ref{eqa2}) and hereafter the summation convention is used. Note that the function $D$ of Eq.~(\ref{eqa2}) is the same as the denominator function of Eq.~(\ref{eq17b}) for $\psi=0$.

As the one-loop contribution to the free energy is $\left(2\beta\right)^{-1}{\rm Tr}\ln \left(\beta S_2\right)$, the one-loop contribution to the dielectric constant is given by
\begin{align}
\label{eqa8}
\eps_1=-&\frac{1}{2\beta}\int \D^3r' \,\frac{\delta^2 {\rm Tr} \ln \left(\beta S_2\right)}{\delta E_z(\vecr)\delta E_z(\vecr')}.
\end{align}
For isotropic systems such as the one discussed here, the $x$ and $y$ Cartesian components of the electric field could have equivalently been used instead of the $z$ component.

Expanding the logarithm of the operator $S_2$ to second order in $E$ yields
\begin{align}
\label{eqa3}
\frac{\delta^2 {\rm Tr} \ln \left(\beta S_2\right)}{\delta E_i(\vecr)\delta E_j(\vecr')}\left(\vecr,\vecr'\right)&=G\left(\vecr,\vecr'\right)S_{2}^{(ij)}\left(\vecr,\vecr'\right)\nonumber\\
&-G^2\left(\vecr,\vecr'\right)\left[S_2^{(i)} S_2 ^{(j)}\right]\left(\vecr,\vecr'\right)\delta_{ij},
\end{align}
where $G=S_2^{-1}$ is the Green's function, defined in Eq.~(\ref{eq22}), and we have defined $S_2^{(i)}=\partial S_2/\partial E_i$  and $S_2^{(ij)}=\partial S_2/\partial E_i\partial E_j$, evaluated at $E=0$. Hereafter, such upper indices refer only to derivatives with respect to the Cartesian components of the electric field. Note that the dependence  of $S_2$ on $E$ is via integrals over the solid angle of the form
\be
\label{eqa4}
g(E)=\frac{1}{4\pi}\int\D^2\Omega\,\e^{-\beta p_k E_k},
\ee
and via derivatives of $g(E)$ with respect to different components of $E$. For example, the two integrals appearing in $h$ [Eq.~(\ref{eqa2})], are proportional to the second-order derivative $g^{(lm)}$ and the product of first-order derivatives, $g^{(l)} g^{(m)}$. For $E=0$, such integrals have radial symmetry. Therefore, any odd derivative of $g(E)$, containing an odd number of vectors $n_i$ in the integrand, vanishes. This is a key feature incorporated in our calculation.

Due to the symmetry argument above, the terms $S_2^{(i)}$ and $S_2^{(j)}$ in Eq.~(\ref{eqa3}) vanish. The second-order derivative $S_2^{(ij)}$ can be simplified by symmetry, according to
\be
\label{eqa5}
\left(\frac{h}{D}\right)^{(ij)}=\frac{h^{(ij)}}{D}+h\left(\frac{1}{D}\right)^{(ij)}.
\ee
Carrying out the derivatives in Eq.~(\ref{eqa5}) leads to
\begin{align}
\label{eqa6}
h^{(ij)}&=\left(\beta e b^2\right)^2\left[\left(-\int \frac{\D^2\Omega}{4\pi} \, \hat{n}_i \hat{n}_j \hat{n}_l \hat{n}_m\right.\right.\nonumber\\ \nonumber \\
&+\left.\left.\frac{1}{9}\frac{\delta_{il}\delta_{jm}+\delta_{im}\delta_{jl}}{1+2\Lambda}\right)\partial_m\delta\left(\vecr-\vecr'\right)\right],\nonumber\\
\left(\frac{1}{D}\right)^{(ij)}&=-\frac{\left(\beta e b\right)^2}{3\left(1+2\Lambda\right)^2}\,\delta_{ij},
\end{align}
where we have used the fact that $g^{(ij)}=\delta_{ij}/3$. Substituting Eqs.~(\ref{eqa5}) and (\ref{eqa6}) in Eq.~(\ref{eqa3}), and performing integration by parts leads to
\begin{widetext}
\begin{align}
\label{eqa7}
&\frac{\delta^2 {\rm Tr} \ln \left(\beta S_2\right)}{\delta E_i(\vecr)\delta E_j(\vecr')}=\frac{\left(4\pi\right)^2 \beta l_0^2 b^2 \eps_0}{a^3\left(1+2\Lambda\right)}\delta\left(\vecr-\vecr'\right)\nonumber\\ \nonumber\\
&\times\left[\left(-\int \frac{\D^2\Omega}{4\pi} \, \hat{n}_i \hat{n}_j \hat{n}_l \hat{n}_m+\frac{1}{9}\frac{\delta_{il}\delta_{jm}+\delta_{im}\delta_{jl}}{1+2\Lambda}\right)b^2\partial_l\partial_m G\left(\vecr,\vecr'\right)\right.
 \left.-\frac{1}{3\left(1+2\Lambda\right)}\delta_{ij}\left(2\Lambda G\left(\vecr,\vecr'\right)-\frac{1}{3}b^2\partial_m\partial_m G\left(\vecr,\vecr'\right)\right)\right].
\end{align}
\end{widetext}

As a result from the integration over the Dirac delta function in Eq.~(\ref{eqa7}), the Green's function and its derivatives are evaluated at $\vecr-\vecr'=0$. The divergence of $G(0)$ is avoided by introducing a cutoff wavelength, as is explained before Eq.~(\ref{eq26}). In particular, second derivatives of the form $\partial_l \partial_m G(0)$ are given by
\begin{align}
\label{eqa9}
\partial_l\partial_m G(0)=-\frac{1}{\emf}\int_{|\veck|<\kmax}^{}\frac{\D^3k}{\left(2\pi\right)^3}\frac{k_l k_m}{k^2+\kmf^2}.
\end{align}
The above integral vanishes unless $l=m$, due to the radial symmetry of the integrand. Therefore, $\partial_l \partial_m G(0)=\delta_{lm}\partial_k \partial_k G(0)/3$, leading to the final result
\begin{align}
\label{eqa10}
\frac{\eps_1}{\eps_0}=\frac{\delta}{\left(1+2\Lambda\right)^2}\left(4\pi l_0\Lambda G(0)-\frac{1-3\Lambda}{3} a^3\delta\nabla^2G(0)\right),
\end{align}
which is used in Eq.~(\ref{eq24}).

\newpage

\end{document}